\documentclass[12pt]{article}
\usepackage{amsmath,amssymb}
\usepackage{graphicx}
\newcommand{\eqdef}{\stackrel{\text{def}}{=}}

\newcommand{\ignore}[1]{}
\numberwithin{equation}{section}
\newcommand{\Romannumeral}[1]{\uppercase\expandafter{\romannumeral#1}}

\allowdisplaybreaks[4]

\begin{document}

\baselineskip=20pt

\newcommand{\preprint}{
\vspace*{-20mm}\begin{flushleft} Version: 0  February 14 2014 \end{flushleft}
}
\newcommand{\Title}[1]{{\baselineskip=26pt
   \begin{center} \Large \bf #1 \\ \ \\ \end{center}}}
\newcommand{\Author}{\begin{center}
   \large \bf  Khosrow Chadan${}^1$, Reido Kobayashi${}^2$ and
   Ryu Sasaki${}^{3}$ \end{center}}
\newcommand{\Address}{\begin{center}
      ${}^1$ Laboratoire de Physique Th\'{e}orique%
      \footnote[1]{Unit\'{e} Mixe de Recherche UMR 8627-CRNS}\\
      Universit\'{e} de Paris XI, B\^{a}timent 210, 91405 Orsay Cedex, France\\
      ${}^2$ Department of Mathematics\\
      Tokyo University of Science, Noda, Chiba 278-8510, Japan\\
      ${}^{3}$ Department of Physics\\  
      Shinshu University, Matsumoto 390-8621, Japan
    \end{center}}
\newcommand{\Accepted}[1]{\begin{center}
   {\large \sf #1}\\ \vspace{1mm}{\small \sf Accepted for Publication}
   \end{center}}

\preprint
\thispagestyle{empty}

\Title{Composition of Two  Potentials}                     %

\Author

\Address
\vspace{1cm}

\begin{abstract}
Given two potentials $V_0$ and $V_1$ together with a {\em certain nodeless solution\/} 
$\varphi_0$ of $V_0$, we form a composition of these two potentials.
If $V_1$ is exactly solvable, the composition is exactly solvable, too.
By combining various solvable potentials in one-dimensional quantum mechanics,
a huge variety of solvable compositions can be made.
\end{abstract}

\section{Introduction}
\label{intro}

Some years ago, two of the present authors developed a method of composing two 
potentials $V_0$ and $V_1$ in a special setting \cite{CK1}.
The potentials were spherically symmetric and of short range and were explicitly solvable 
at zero energy. The starting potential $V_0$ should sustain no bound states.
Then the composed potential was explicitly solvable at zero energy, too.
The main motivation was applications to quantum mechanical scattering problems 
\cite{CS} and the radial Sch"oinger equations.

In the present paper, we show that the composition of two potentials $V_0$ and $V_1$
in generic one dimensional quantum mechanics can be defined in a much more general setting.
The only requirement is that $V_0$ should have a {\em nodeless solution\/} 
satisfying certain boundary conditions.
If $V_1$ is exactly solvable, so is the composed potential $V_C$.
If $V_1$ has also a nodeless solution satisfying boundary conditions, a further composition
with another potential $V_2$ can be made, as emphasised in the earlier papers \cite{CK1}.
We provide several examples of $V_0$, all exactly solvable, 
together with an infinite number of nodeless solutions satisfying the boundary conditions.
These nodeless solutions are called {\em virtual state wavefunctions\/} \cite{os25,os28},
which have played an important role in the construction of various rational deformations of 
exactly solvable potentials and the {\em multi-indexed orthogonal polynomials\/}
\cite{os25}--\cite{hos} as the main part of the eigenfunctions of the deformed solvable systems.

This paper is organised as follows.
In section \ref{sec:comp}, we formulate the composition of two potentials in the most general setting.
Derivation is quite elementary. Section \ref{exam} provides explicit examples of nodeless
solutions (virtual state wavefunctions) belonging to the well-known 
solvable potentials \cite{infhul,susyqm};
the radial oscillator potential in \S\ref{rado}, P\"oschl-Teller potential in \S\ref{PT},
hyperbolic P\"oschl-Teller potential in \S\ref{hPT}, Rosen-Morse potential in \S\ref{RM}
and Eckart potential in \S\ref{eck}.  The final section is for a short summary and comments.

\section{General Composition}
\label{sec:comp}
The main ingredients of the present theory  are two quantum mechanical systems
in one dimension with smooth potentials $V_0$ and $V_1$:
\begin{align}
\text{System}\ 0:\qquad &\mathcal{H}_0=-\frac{d^2}{dx^2}+V_0(x),\qquad \quad a<x<b,\\
                  &\mathcal{H}_0\varphi_0(x)=\tilde{\mathcal E}_0\varphi_0(x),\\
\text{System}\ 1:\qquad &\mathcal{H}_1=-\frac{d^2}{dx^2}+V_1(x),\qquad \quad c<x<d,\\
                  &\mathcal{H}_1\phi_m(x)={\mathcal E}_m\phi_m(x),\qquad \quad m=1,\ldots,                  
\end{align}
Since they are both quantum mechanical systems, a finite boundary can only be realised by
a strong repulsive singularity.
Otherwise, the wave functions would leak beyond that boundary and the quantum system
is ill-defined.
The System 0 needs not be solvable. 
The only requirement is that it has a {\em nodeless real solution\/} $\varphi_0(x)$ with proper
boundary conditions (A) and (B), to be specified later.
The System 0 may or may not have some discrete eigenstates, either finitely many or infinite in number.
The System 1 is a general quantum system.
Nothing special is required of $V_1(x)$ for the construction of the composition
of the two potentials $V_0$ and $V_1$.

It is elementary to show that the following function\begin{equation}
\chi_0(x)\eqdef\varphi_0(x)\left(\alpha+\int_{x}^{b}\frac{dt}{\varphi_0^2(t)}\right),\quad \alpha\eqdef\frac1{d-c},
\end{equation}
 is another solution of $\mathcal{H}_0$
with the same energy $\tilde{\mathcal E}_0$:
\begin{equation}
\mathcal{H}_0\chi_0(x)=\tilde{\mathcal E}_0\chi_0(x).
\end{equation}
Let us introduce a mapping function $\psi_0(x)$, $a<x<b$,
\begin{equation}
\psi_0(x)\eqdef c+\frac{\varphi_0(x)}{\chi_0(x)}=c+\frac1{\alpha+\int_{x}^{b}\frac{dt}{\varphi_0^2(t)}},
\label{mappsi}
\end{equation}
which is monotonously increasing
\begin{equation*}
\frac{d\psi_0(x)}{dx}=\frac1{\chi_0^2(x)}>0.
\end{equation*}
We assume the following boundary conditions on $\varphi_0(x)$:
\begin{align*}
\hspace{43mm} x\to a\qquad &\int_{x}^{b}\frac{dt}{\varphi_0^2(t)}\to \infty \hspace{66mm} (\text{A}),\\
\hspace{43mm}  x\to b\qquad &\int_{x}^{b}\frac{dt}{\varphi_0^2(t)}\to 0 \hspace{68mm}  (\text{B}),
\end{align*}
which mean 
\begin{equation}
\psi_0(a)=c,\quad \psi_0(b)=c+\frac1\alpha=d.
\end{equation}
Thus $\psi_0(x)$ is a 1:1 onto mapping from $(a,b)$ to $(c,d)$.

The composition of two potentials,
$V_0$ and $V_1$  is achieved by $\chi_0(x)$ and $\psi_0(x)$:
\begin{equation}
\mathcal{H}_C=-\frac{d^2}{dx^2}+V_C(x),\quad 
V_C(x)\eqdef V_0(x)-\tilde{\mathcal E}_0+\frac1{\chi_0^4(x)}V_1(\psi_0(x)),\quad a<x<b.
\end{equation}
With a solution $\phi_m(x)$ of $\mathcal{H}_1$, let us define
\begin{equation}
\phi_m^C(x)\eqdef \chi_0(x)\phi_m(\psi_0(x)), \quad m=1,\ldots, .
\end{equation}
It is elementary to show that $\phi_m^C(x)$ satisfies
\begin{equation}
\mathcal{H}_C\phi_m^C(x)=\frac{\mathcal{E}_m}{\chi_0^4(x)}\phi_m^C(x),\quad m=1,\ldots.
\end{equation}
If $\{\phi_m(x)\}$ are eigenfunctions of $\mathcal{H}_1$:
\begin{equation}
\int_c^d\!\!\phi_m(x)\phi_n(x)dx=h_m\delta_{m\,n},\quad h_m>0,\quad m,n=1,\ldots,
\end{equation}
then $\{\phi_m^C(x)\}$ are orthogonal with each other with respect to 
the weight function $\frac1{\chi_0^4(x)}$:
\begin{align} 
 \int_a^b\!\!\phi_m^C(x)\phi_n^C(x)\frac{dx}{\chi_0^4(x)}&=\int_a^b\!\!\phi_m(\psi_0(x))\phi_n(\psi_0(x))\frac{dx}{\chi_0^2(x)}\\
 &=\int_c^d\!\!\phi_m(\psi_0)\phi_n(\psi_0)d\psi_0=h_m\delta_{m\,n},\quad m,n=1,\ldots,.
\end{align}
If $\mathcal{H}_1$ is exactly solvable, so is the composed Hamiltonian $\mathcal{H}_C$.

As stressed in \cite{CK1}, further compositions are straightforward, so long as the
proper nodeless solutions of can be found.  Suppose we have another Hamiltonian system 
$\mathcal{H}_2$ with a potential $V_2$
\begin{align*}
\text{System}\ 2:\qquad &\mathcal{H}_2=-\frac{d^2}{dx^2}+V_2(x),\qquad \qquad e<x<f,\\
                  &\mathcal{H}_2\Phi_m(x)={E}_m\Phi_m(x),\qquad \quad m=1,\ldots,                  
\end{align*}
and a nodeless solution $\varphi_1$ of $\mathcal{H}_1$ with proper boundary conditions. 
The next step is essentially the same as before:
\begin{align*}
\mathcal{H}_1\varphi_1(x)=\tilde{\mathcal E}_1\varphi_1(x),\quad 
\chi_1(x)&\eqdef \varphi_1(x)\left(\alpha_1+\int_{x}^{d}
\!\frac{dt}{\varphi_1^2(t)}\right),\quad \alpha_1=(f-e)^{-1},\\
\mathcal{H}_1\chi_1(x)=\tilde{\mathcal E}_1\chi_1(x),\quad 
\psi_1(x)&\eqdef e+\frac{\varphi_1(x)}{\chi_1(x)}=e+\frac1{\alpha_1+\int_{x}^{d}
\!\frac{dt}{\varphi_1^2(t)}},\quad
 \frac{d\psi_1(x)}{dx}=\frac1{\chi_1^2(x)}>0.
\end{align*}
The second composed Hamiltonian is
\begin{align*}
\mathcal{H}_{C2}&=-\frac{d^2}{dx^2}+V_{C2}(x),\qquad \qquad a<x<b,\\
V_{C2}(x)&\eqdef V_0(x)-\tilde{\mathcal E}_0+\frac1{\chi_0^4(x)}\left(V_1(\psi_0(x))-\tilde{\mathcal E}_1\right)+\frac1{\chi_0^4(x)\chi_1^4(\psi_0(x))}
V_2\bigl(\psi_1(\psi_0(x))\bigr), 
\end{align*}
with its solutions
\begin{align*}
\Phi_m^{C2}(x)&\eqdef \chi_0(x)\chi_1(\psi_0(x))\Phi_m\bigl(\psi_1(\psi_0(x))\bigr),\quad m=1,\ldots,\\
\mathcal{H}_{C2}\Phi_m^{C2}(x)&=\frac{E_m}{\chi_0^4(x)\chi_1^4(\psi_0(x))}\Phi_m^{C2}(x).
\end{align*}
The composition processes can go on indefinitely.

\section{Explicit Examples}
\label{exam}

Several explicit examples of the compositions were demonstrated in \cite{CK1} in connection with
certain scattering problems \cite{CS}.
For example, given two spherically symmetric and short range potentials for which
the radial Schr\"odinger equations can be solved at zero energy, the composition can 
also be solved at zero energy.
Here we give several explicit examples of the System 0, together with the proper
{\em nodeless solutions\/} $\varphi_0(x)$. Those examples presented in \cite{CK1}
will not be re-listed here.
Although System 0 needs not be solvable, most examples listed below are, in fact,
exactly solvable. 
The reason  is quite trivial.
The proper nodeless solutions can be most easily found when the system is solvable.
The nodeless solutions listed below are called {\em virtual state wavefunctions\/}
\cite{os25,os28}.
Their energies are {\em below the ground state energy\/} and they together with their inverses are 
{\em square non-integrable\/}.
Thus the conditions (A) and (B) are satisfied. In most cases, they are obtained from
the eigenfunctions by discrete symmetry transformations.
They have played an essential role in the rational deformations of solvable potentials,
creating the multi-indexed and exceptional orthogonal polynomials \cite{os25}, 
\cite{gomez3}--\cite{hos}, which are new species of orthogonal polynomials satisfying second order differential
equations but not the three term recurrence relations \cite{boch}.

See for example \cite{infhul,susyqm} for a general review of exactly solvable potentials.
We follow \cite{infhul} for the naming of solvable potentials.
Detailed information of these systems, the symmetry, eigenfunctions, 
various virtual state wave functions and their generalisation can be found in \cite{os25,os28}.
The first two examples have infinitely many discrete eigenstates.

\subsection{Radial oscillator}
\label{rado}
The radial oscillator potential is
\begin{equation}
V_0(x)\eqdef x^2+\frac{g(g-1)}{x^2}, \quad 0<x<\infty,\quad g>1/2.
\end{equation}
The lower boundary $x=0$  is a regular singularity with the characteristic exponents 
$g$ and $1-g$. 
In terms of a discrete symmetry transformation $x\to ix$, the eigenfunction with degree v is mapped to
a virtual state wavefunction
\begin{equation}
\varphi_{0,\text{v}}(x)\eqdef e^{x^2/2}x^gL_\text{v}^{(g-1/2)}(-x^2),\quad
\tilde{\mathcal E}_{0,\text{v}}=-4\text{v}-2g+3,
\quad \text{v}=0,1,\ldots,
\end{equation}
in which $L_n^{(\alpha)}(x)$ is the Laguerre polynomial of degree $n$  in $x$. 
The nodelessness of $\varphi_{0,\text{v}}(x)$ is obvious, since all the zeros of the Laguerre polynomial $L_n^{(\alpha)}(x)$ are positive. 
The boundary conditions (A) and (B) are satisfied.
The Wronskian%
\footnote{$\text{W}[f_1,\cdots,f_n](x)\eqdef\det\bigl(\partial_x^{j-1}
f_k(x)\bigr)_{1\le j,k\le n}$.} 
of these virtual state wave functions
\begin{equation}
\text{W}[\varphi_{0,\text{v}_1},\ldots, \varphi_{0,\text{v}_M}](x)
\end{equation}
can also be used for compositions, as demonstrated for the multi-indexed orthogonal 
polynomials \cite{os25}.
The situation is the same for all the other examples below.
The radial oscillator itself provides an infinite number of different compositions.

\subsection{P\"oschl-Teller}
\label{PT}
The P\"oschl-Teller potential is
\begin{equation}
V_0(x)=\frac{g(g-1)}{\sin^2x}+\frac{h(h-1)}{\cos^2x},\quad 0<x<\frac{\pi}2,\quad g,h>1/2.
\end{equation}
The lower boundary $x=0$  is a regular singularity with the characteristic exponents 
$g$ and $1-g$. The upper boundary $x=\frac{\pi}2$ is also a regular singularity with the characteristic exponents 
$h$ and $1-h$.
In terms of a discrete symmetry transformation $h\to 1-h$, 
the eigenfunction with lower degree v is mapped to
a virtual state wavefunction
\begin{align}
\varphi_{0,\text{v}}(x)&\eqdef (\sin x)^g(\cos x)^{1-h}
P_\text{v}^{(g-1/2,1/2-h)}(\cos2x),\\
 \tilde{\mathcal E}_{0,\text{v}}&\eqdef(g-h+1+2\text{v})^2,\qquad \qquad \text{v}=0,1,\ldots,[h-1/2]'.
\end{align}
Here $P_n^{(\alpha,\beta)}(x)$ is the Jacobi polynomial of degree $n$ in $x$  and $[a]'$ denotes the greatest integer less than $a$.
It is easy to see that the boundary conditions (A) and (B) are satisfied.
The expansion of the Jacobi polynomial can be used to demonstrate the nodelessness. (See 
(3.2)  of \cite{os21}.)

\subsection{Hyperbolic P\"oschl-Teller}
\label{hPT}
The hyperbolic P\"oschl-Teller potential 
\begin{equation}
V_0(x)=\frac{g(g-1)}{\sinh^2x}-\frac{h(h+1)}{\cosh^2x},\quad 0<x<\infty,\quad h>g>1/2,
\end{equation}
has $[(h-g)/2]'+1$ discrete eigenstates.
The lower boundary $x=0$  is a regular singularity with the characteristic exponents 
$g$ and $1-g$. 
By the discrete symmetry transformation $h\to -(h+1)$, the eigenfunction with degree v
is mapped to a virtual state wave function
\begin{align}
\varphi_{0,\text{v}}(x)&\eqdef(\sinh x)^g(\cosh x)^{h+1}
  P_{\text{v}}^{(g-1/2,h+1/2)}(\cosh2x),\\
 \tilde{\mathcal E}_{0,\text{v}}&\eqdef-(h-g)^2-(2\text{v}+2g+1)(2\text{v}+2h+1),\quad 
  \text{v}=0,1,\ldots,[(h-g)/2]'.
\end{align}
It is easy to see that the boundary conditions (A) and (B) are satisfied.
The expansion of the Jacobi polynomial can be used to demonstrate the nodelessness.

The {\em overshoot eigenfunctions\/} \cite{os28} also provides the proper nodeless solutions.
They have exactly the same form as the
eigenfunctions but their degrees are much higher than that  of the highest eigenstate
so that their energies are lower than the groundstate energy:
\begin{align}
\varphi_{0,\text{v}}(x)&\eqdef(\sinh x)^g(\cosh x)^{-h}
  P_{\text{v}}^{(g-1/2,-h-1/2)}(\cosh2x),\\
 \tilde{\mathcal E}_{0,\text{v}}&\eqdef-(h-g-2\text{v})^2,\quad 
  \text{v}>h-g.
\end{align}
It is easy to see that the boundary conditions (A) and (B) are satisfied.
The expansion of the Jacobi polynomial can be used to demonstrate the nodelessness.

\subsection{Rosen-Morse}
\label{RM}
This potential is
\begin{equation}
V_0(x)\eqdef -\frac{h(h+1)}{\cosh^2x}+2\mu\tanh x,\quad   -\infty<x<\infty,\quad h>\sqrt{\mu}>0.
\end{equation}
The system has finitely many discrete eigenstates
$[h-\sqrt{\mu}\,]'+1$. The overshoot eigenfunctions $h<\text{v}<h+\frac{\mu}{h}$ 
provides proper nodeless solutions:
\begin{align}
\varphi_{0,\text{v}}(x)&\eqdef e^{-\frac{\mu}{h-\text{v}}x}(\cosh x)^{-h+\text{v}}
  P_{\text{v}}^{(\alpha_{\text{v}},\beta_{\text{v}})}(\tanh x),\\
  \alpha_\text{v}&\eqdef h-\text{v}+\frac{\mu}{h-\text{v}},\ \ \beta_\text{v}\eqdef h-\text{v}-\frac{\mu}{h-\text{v}},\\
 \tilde{\mathcal E}_{0,\text{v}}&\eqdef-(h-\text{v})^2-\frac{\mu^2}{(h-\text{v})^2},\quad 
 h< \text{v}<h+\frac{\mu}{h}.
\end{align}
It is easy to see that the boundary conditions (A) and (B) are satisfied.
The expansion of the Jacobi polynomial can be used to demonstrate the nodelessness.

\subsection{Eckart}
\label{eck}
This potential problem is also called Kepler problem in hyperbolic space:
\begin{equation}
V_0(x)\eqdef \frac{g(g-1)}{\sinh^2x}
  -2\mu\coth x, \quad
  0<x<\infty,\quad \sqrt{\mu}>g>\frac12.
\end{equation}
It has a finite number of discrete eigenstates, $[\sqrt{\mu}-g]'+1$.
The lower boundary $x=0$  is a regular singularity with the characteristic exponents 
$g$ and $1-g$.
The overshoot eigenfunctions $\text{v}>\frac{\mu}{g}-g$ 
provide proper nodeless solutions:
\begin{align}
\varphi_{0,\text{v}}(x)&\eqdef e^{-\frac{\mu}{g+\text{v}}x}(\sinh x)^{g+\text{v}}
  P_{\text{v}}^{(\alpha_{\text{v}},\beta_{\text{v}})}(\coth x),\\
  \alpha_\text{v}&\eqdef -g-\text{v}+\frac{\mu}{g+\text{v}},\ \ \beta_\text{v}\eqdef -g-\text{v}-\frac{\mu}{g+\text{v}},\\
 \tilde{\mathcal E}_{0,\text{v}}&\eqdef-(g+\text{v})^2-\frac{\mu^2}{(g+\text{v})^2},\quad 
 \text{v}>\frac{\mu}{g}-g.
\end{align}
It is easy to see that the boundary conditions (A) and (B) are satisfied.
The expansion of the Jacobi polynomial can be used to demonstrate the nodelessness.

\section{Summary and comments}
\label{summary}

General prescriptions for composing two potentials $V_0$ and $V_1$ in one dimensional quantum mechanics 
are presented by extending the original work in \cite{CK1}.
The virtual state wavefunctions \cite{os25} play an important role, as they have played in 
the rational deformations of solvable potentials \cite{os25}--\cite{hos}.
The form of the mapping function $\psi_0(x)$ \eqref{mappsi} is reminiscent of the
Abraham-Moses transformations \cite{A-M}, \cite{pursey}, \cite{os31}, or the so-called
binary Darboux transformations \cite{matv}.
The virtual state wavefunctions also play important roles in Abraham-Moses transformations \cite{os31}.
\section*{Acknowledgements}

 R.\,S. is supported in part by Grant-in-Aid for Scientific Research
from the Ministry of Education, Culture, Sports, Science and Technology
(MEXT),  No.22540186.


\bigskip

\bigskip

\appendix{\Huge{\textbf{{Appendix }}}}

%
%
%



\end{document}